\documentstyle[preprint,eqsecnum,aps]{revtex}
\begin{document}
\draft
\title{Intrinsic surface depression of the order parameter under mixed ({\it s + id)%
}-wave pair symmetry and its effect on the critical current of high-Tc SIS
Josephson junctions}
\author{G.A. Ummarino$^a$, R.S. Gonnelli$^a$, C. Bravi$^a$ and V.A. Stepanov$^b$}
\address{$^a$ INFM-Dipartimento di Fisica, Politecnico di Torino, Torino, Italy\\
$^b$ P.N. Lebedev Physical Institute, Russian Academy of Sciences, Moscow,
Russia}
\maketitle

\begin{abstract}
An intrinsic gap depression at the Superconductor-Insulator interface due to
the very short value of the coherence length in High-Tc Superconductors
[HTSs] is considered, in the framework of a mixed ({\it s+id})-wave pair
symmetry for the order parameter ranging from pure {\it s} to pure {\it d}%
-wave. This gap depression acts as the main physical agent causing the
relevant reduction of $I_c\left( T\right) R_n(T)$ values with respect to
BCS\ expectations in HTS SIS Josephson junctions. Good agreement with
various experimental data is obtained with both pure {\it s}-wave and pure 
{\it d}-wave symmetries of the order parameter, but with amounts of gap
depression depending on the pair symmetry adopted. Regardless of the pair
symmetry considered, these results prove the importance of the surface
order-parameter depression in the correct interpretation of the $%
I_c(T)R_n(T) $ data in HTS SIS junctions. In a case of planar YBCO-based
junction the use of the de Gennes condition allowed us to tentatively obtain
an upper limit for the amount of {\it d}-wave present in the gap of YBCO.
\end{abstract}

\pacs{PACS numbers: 74.50.+r; 74.70.Vy\\
KEY WORDS: High-Tc superconductors; Order parameter; Josephson junctions;
Josephson critical current; Pair symmetry.}

As a consequence of the very short coherence length $\xi $ of high-Tc
superconductors [HTSs], an intrinsic depression of the order-parameter at
Superconductor-Insulator [S-I] interfaces arises \cite{deutscher87}, which
can represent one of the main reason for the reduced $I_c(T)R_n(T)$ values
in HTS\ SIS \cite{mannhart91,gonn95} and SIS' Josephson junctions \cite
{gonn96a}.

Also pair symmetry is expected to contribute to such experimental results,
and, in the present paper, a model is developed by us which takes into
consideration pure {\it s}-wave, pure {\it d}-wave and mixed $(s+id)$-wave
pair symmetry together with a suitable gap depression, extending and
generalizing our previous model developed for $T>0$ only in the pure {\it s}%
-wave case \cite{gonn95,gonn96a}.

By making use of the space-dependent expression of the gap $\Delta \left(
x,T\right) $= $\Delta \left( T\right) $ $\tanh \left[ \left( x+x_0\right)
/\left( \sqrt{2}\xi \left( T\right) \right) \right] \cdot $ $\theta \left(
x-w\right) $ which may be derived from Ginzburg-Landau equations, $%
I_c(T)R_n(T)$ values in SIS and SIS' Josephson junctions [JJs] may be
evaluated. Here $2w$ is the thickness of the insulating layer and $x_0$
accounts for the spatial slope of $\Delta \left( x,T\right) $ near interfaces%
\cite{gonn96a}. As in Ref. \cite{sigrist91} we assume that the {\it d}-wave
component of the order parameter $\Delta _d$ has the same spatial behaviour
as the {\it s}-wave component $\Delta _s$ and, moreover, we neglect the
fourth-order terms in the Ginzburg-Landau equations for the $(s+id)$-wave
pair symmetry. We imagine to ''section''\ the gap $\Delta \left( x,T\right) $
into independent channels $\delta \Delta _i$ \cite{gonn95,gonn96a}, each of
them giving rise to a parallel contribution $\delta I_{c_i}\left( T\right) $
and $\delta G_{n_i}\left( T\right) $ to the total critical current and
normal conductance, respectively. The physical situation just described is
depicted in Fig. 1 where our model for a HTS SIS Josephson junction is
shown. By summing and averaging over all the parallel contributions, that in
general differ since each ''slice'' corresponds to a different thickness of
the barrier (see Fig. 1), we obtain:

\[
I_c\left( T\right) \simeq \frac 1{\Delta \left( T\right) }%
\sum\limits_i\left[ I_{c_i}\left( T\right) \cdot \delta \Delta _i\left(
x,T\right) \right] 
\]

\[
G_n\left( T\right) \simeq \frac 1{\Delta \left( T\right) }%
\sum\limits_i\left[ G_{n_i}\left( T\right) \cdot \delta \Delta _i\left(
x,T\right) \right] 
\]

In the continuous limit these expressions become:

\[
I_c\left( T\right) =\frac 1{\Delta \left( T\right) }\int\nolimits_0^\infty
I_c\left( x,T\right) \frac{\partial \Delta \left( x,T\right) }{\partial \ x}%
dx 
\]

\[
G_n\left( T\right) =\frac 1{\Delta \left( T\right) }\int\nolimits_0^\infty
G_n\left( x,T\right) \frac{\partial \Delta \left( x,T\right) }{\partial \ x}%
dx 
\]

and

\[
\left[ G_n\left( x,T\right) \right] ^{-1}=R_n\left( x,T\right) =R_0\left(
T\right) \cdot 10^{2\left( \frac xw-1\right) } 
\]

where $R_0$ is a parameter representing the resistance of the channel at $%
x=w $, which gets eliminated after performing the integral for $G_n\left(
T\right) $, and where the space dependence of $R_n\left( x,T\right) $ has
been heuristically derived under global conditions of consistency \cite
{gonn95}. The expression for the spatial dependence of $I_c\left( T\right) $
in mixed ({\it s+id})-wave symmetry is derived by introducing the spatial
dependencies of $\Delta _s$ and $\Delta _d$ in the results of Ref. \cite
{xu94} :

\begin{equation}
I_c\left( x,T\right) =\frac T{e\pi \cdot R_n(x,T)}\sum_{l=1}^\infty \frac{%
4\Delta _s^2\left( x,T\right) +\Delta _d^2\left( x,T\right) }{\omega
_l^2+\Delta _s^2\left( x,T\right) +\Delta _d^2\left( x,T\right) }\left[
K\left( \frac{\Delta _d\left( x,T\right) }{\sqrt{\omega _l^2+\Delta
_s^2\left( x,T\right) +\Delta _d^2\left( x,T\right) }}\right) \right] ^2 
\eqnum{1}  \label{Eq:1}
\end{equation}

where $\omega _l=\left( 2l-1\right) \pi k_BT$ is the Matsubara frequency and 
$K$ is the complete elliptic integral of the first kind.

After performing rather cumbersome calculations, the general expression for
the temperature dependent critical voltage $I_c(T)R_n(T)$ in the case of
SIS\ JJs, in mixed ({\it s+id})-wave symmetry and in presence of surface
depression of the order parameter is given by:

\begin{equation}
\left[ I_c(T)R_n(T)\right] _{SIS}^{\left( s+id\right) +dep.gap}=\frac{4k_BT}{%
e\pi }\cdot A\left( T\right) \cdot \sum\limits_{l=1}^\infty \left[ B\left(
l,T\right) \cdot K_\gamma ^2\left( l,T\right) +\int\limits_{\gamma \left(
T\right) }^1C\left( l,T,z\right) \cdot K_z^2\left( l,T,z\right) \cdot
dz\right]  \eqnum{2}  \label{Eq:2}
\end{equation}

where

\[
A\left( T\right) =\frac 1{\gamma \left( T\right) \left( \frac{1-\gamma
\left( T\right) }{1+\gamma \left( T\right) }\right) ^{\vartheta \left(
T\right) }+\int\limits_{\gamma \left( T\right) }^1dz\left( \frac{1-z\left(
T\right) }{1+z\left( T\right) }\right) ^{\vartheta \left( T\right) }} 
\]

\[
B\left( l,T\right) =\frac{\gamma ^3\left( T\right) \left( \frac{1-\gamma
\left( T\right) }{1+\gamma \left( T\right) }\right) ^{\vartheta \left(
T\right) }\left[ 4\Delta _s^2\left( T\right) +\Delta _d^2\left( T\right)
\right] }{\omega _l^2+\gamma ^2\left( T\right) \left( \Delta _s^2\left(
T\right) +\Delta _d^2\left( T\right) \right) } 
\]

\[
C\left( l,T,z\right) =\frac{z^2\left( T\right) \left( \frac{1-z\left(
T\right) }{1+z\left( T\right) }\right) ^{\vartheta \left( T\right) }\left[
4\Delta _s^2\left( T\right) +\Delta _d^2\left( T\right) \right] }{\omega
_l^2+z^2\left( T\right) \left( \Delta _s^2\left( T\right) +\Delta _d^2\left(
T\right) \right) } 
\]

\[
K_\gamma \left( l,T\right) =K\left( \frac{\Delta _d\left( T\right) \ \gamma
\left( T\right) }{\sqrt{\omega _l^2+\gamma ^2\left( T\right) \left( \Delta
_s^2\left( T\right) +\Delta _d^2\left( T\right) \right) }}\right) 
\]

\[
K_z\left( l,T,z\right) =K\left( \frac{\Delta _d\left( T\right) \ z\left(
T\right) }{\sqrt{\omega _l^2+z^2\left( T\right) \left( \Delta _s^2\left(
T\right) +\Delta _d^2\left( T\right) \right) }}\right) 
\]

and $z\left( x,T\right) =\tanh \left( \frac{x+x_0}{\xi \left( T\right) \sqrt{%
2}}\right) ,$ $\gamma \left( T\right) =\tanh \left( \frac{w+x_0}{\xi \left(
T\right) \sqrt{2}}\right) ,$ $\vartheta \left( T\right) =\xi \left( T\right)
\left( \frac{\ln \left( 100\right) }{w\sqrt{2}}\right) .$

In the previous expressions we considered $\Delta _s\left( T\right) =\left(
1-\varepsilon \right) \cdot \Delta _{ex}\left( T\right) $ and $\Delta
_d\left( T\right) =\varepsilon \ \Delta _{ex}\left( T\right) $ where $%
\varepsilon $ is the fraction of {\it d}-wave present in the order
parameter, $\Delta _{ex}$ is the experimental value of the gap determined,
for example, in tunneling experiments \cite{zio,cucolo96} and where we have
used for the temperature dependence of $\Delta \left( T\right) $ and $\xi
\left( T\right) $ the standard BCS expressions. In the special case of pure 
{\it s}-wave pair symmetry Eq. (2) reduces to \cite{gonn95}:

\[
\left[ I_c(T)R_n(T)\right] _{SIS}^{pure\ s+dep.\ gap}=\frac{\pi \Delta
_s\left( T\right) }e\frac{\left[ \frac{\gamma ^2}2\left( \frac{1-\gamma }{%
1+\gamma }\right) ^\vartheta \tanh \left( \frac{\gamma \ \Delta _s\left(
T\right) }{2\ k_BT}\right) +\int\nolimits_\gamma ^1z\left( \frac{1-z}{1+z}%
\right) ^\vartheta \tanh \left( \frac{z\ \Delta _s\left( T\right) }{2\ k_BT}%
\right) \ dz\right] }{\gamma \left( \frac{1-\gamma }{1+\gamma }\right)
^\vartheta +\int\nolimits_\gamma ^1\left( \frac{1-z}{1+z}\right) ^\vartheta
dz} 
\]

which tends to the Ambegaokar-Baratoff [AB] model \cite{ambega63} in the
limits valid for a low-T$_c$ superconductor (LTS) (i.e. $x_0\rightarrow
+\infty ,\xi \left( 0\right) \rightarrow +\infty ,x_0/\xi \left( 0\right)
\rightarrow +\infty $):

\[
\lim_{HTS\rightarrow LTS}\left[ I_c(T)R_n(T)\right] _{SIS}^{pure\ s+dep.gap}=%
\frac{\pi \Delta _s\left( T\right) }{2e}\tanh \left( \frac{\Delta _s\left(
T\right) }{2k_BT}\right) . 
\]

When gap depression at S-I\ interfaces is neglected, the result obtained by
Xu et al. \cite{xu94} in the general mixed $(s+id)$-wave case is reproduced: 
\[
\left[ I_c(T)R_n(T)\right] _{SIS}^{s+id}=\frac T{e\pi }\cdot
\sum\limits_{l=1}^\infty \frac{4\Delta _s^2(T)+\Delta _d^2(T)}{\omega
_l^2+\Delta _s^2(T)+\Delta _d^2(T)}\cdot \left[ K\left( \frac{\Delta _d(T)}{%
\sqrt{\omega _l^2+\Delta _s^2(T)+\Delta _d^2(T)}}\right) \right] ^2. 
\]

On the other hand, in the case of pure {\it d}-wave pair symmetry and in
presence of a depression of the order parameter all the components of Eq.
(2) reduce to the analogous expressions for $\Delta _s(T)=0$ (i.e. $%
\varepsilon =1$) : 
\[
\left[ I_c(T)R_n(T)\right] _{SIS}^{pure\text{ }d+dep.gap}=\lim_{\Delta
_s\rightarrow 0}\left[ I_c(T)R_n(T)\right] _{SIS}^{(s+id)+dep.gap}. 
\]

We remark that the free parameters of the model expressed by Eq. (2) in the
most general case are three: $w$, $x_0$ and $\varepsilon $. When the
thickness of the insulating barrier is known (for example in planar SIS\
JJs) the number of free parameters reduces to two. It becomes only one if,
in a planar SIS junction, the amount of {\it d}-wave is also fixed as in the
pure {\it s}-wave case ($\varepsilon =0$) or in the pure {\it d}-wave one ($%
\varepsilon =1$).

A reasonable estimation of the minimum (intrinsic) amount of gap depression
present at S-I interfaces may be obtained by calculating the $x_0$ value -
from now on called $x_0^{dG}$ - that derives from the de Gennes condition at
the S-I interface \cite{degennes66}:

\[
\left[ \frac{d\Delta \left( x\right) }{dx}\right] _{x=w}=\left[ \frac 1b%
\Delta \left( x\right) \right] _{x=w} 
\]

where, in the hypothesis of Ginzburg-Landau behaviour for $\Delta \left(
x\right) $, we have 
\[
\frac 1b=\frac{\sqrt{2}}{\xi \left( 0\right) }\frac 1{\sinh \left( 2\frac{%
w+x_0^{dG}}{\xi \left( 0\right) \sqrt{2}}\right) }. 
\]

The previous expression, together with the approximate value $b\simeq \xi
^2\left( 0\right) /(a-w)$ determined by de Gennes, where $a$ is the range of
the $x$ axis where $\Delta \left( x\right) $ varies appreciably \cite
{degennes66}, permits to calculate $x_0^{dG}$. When, due to a short $\xi
\left( 0\right) $, a large intrinsic gap depression at S-I interfaces is
expected, $\frac{w+x_0^{dG}}{\xi \left( 0\right) \sqrt{2}}<0.4$ so that $%
b\simeq w+x_0^{dG}$. This condition is always satisfied in the HTS junctions
we will describe thereafter. If we impose that $\Delta \left( a,T\right)
/\Delta \left( \infty ,T\right) =0.99,$ considering that $0.99\simeq \tanh
2.6$, a value of $a$ may be yielded so that an estimate for $x_0^{dG}$ is
obtained by means of $x_0^{dG}\simeq 0.6\cdot \xi \left( 0\right) -w$. In
principle, the $x_0$ values determined by fitting the experimental data with
our depressed-gap model have to be comprised between $x_0^{dG}$\ and $x_0$ = 
$-w$ that corresponds to a complete gap depression at the interface i.e. $%
\Delta \left( w\right) $= $0$.

Agreement of such a model with experimental $I_c(T)R_n(T)$ data in HTS SIS
JJs has been tested for the two cases of pure {\it s}-wave and pure {\it d}%
-wave pair symmetry. On the other hand, a fit of the experimental curves by
the complete ({\it s+id})-wave model in absence of indications concerning
the amplitude of the {\it d}-wave component of the gap would have been of
very modest interest due to the numerical impossibility to separate the
contributions of the gap depression $x_0$ and of the {\it d}-wave gap $%
\varepsilon $ to the lowering of $I_c(T)R_n(T)$. As we will see in the
following examples, results are very good with both types of pair symmetry,
but with a different amount of gap depression in each case.

Examples presented in this work include one of our recent $I_c(T)R_n(T)$
behaviours obtained from reproducible nonhysteretic {\it I-V} curves in Bi$%
_2 $Sr$_2$CaCu$_2$O$_{8+x}$ ($T_c=85$ K, $\Delta _{BSCCO}\left( 0\right) =23$
meV, $\xi _{BSCCO}\left( 0\right) =1.67$ nm) single-crystal high-quality
Josephson break junctions (Fig.2). RSJ model \cite{barone82} has been used
to obtain $I_c(T)$ and $R_n(T)$ values as these latter were high enough to
permit to treat the junctions as tunnel SIS ones \cite{gonn95,likharev79}.
Best-fit values for the amount of the gap depression are $\left( x_0\right)
_s=-1.75$ nm and $\left( x_0\right) _d=0.5$ nm, while the thickness of the
barrier, fitted by using the pure {\it s}-wave expression and held constant
also for the pure {\it d}-wave case, is $2w=4.4$ nm. Figure 2 shows the
results of the fit of experimental data by using the depressed-gap model
compared to the standard BCS results (no gap depression) in {\it s}- and 
{\it d}-wave symmetry. As in all the figures from 2 to 5, open symbols are
experimental $I_c(T)R_n(T)$ data, dash-dots and dots represent predictions
obtained considering pure {\it s}-wave (AB model \cite{ambega63}) and pure 
{\it d}-wave pair symmetry {\it without} any gap depression, respectively,
while the results of our model applied to the pure {\it s}-wave and the pure 
{\it d}-wave cases are reported with a dashed and continuous line,
respectively. In all the presented cases the superconducting gap $\Delta (0)$
has been determined from tunneling experiments or from literature while the
coherence length $\xi \left( 0\right) $ was taken from literature.

The depressed-gap model has been tentatively used also in Bi$_2$Sr$_2$CuO$%
_{6+x}$ (Bi-2201) single-crystal break-junctions \cite{gonn96b} where $%
T_c=9.6$ K, $\Delta _{BSCO}\left( 0\right) =3$ meV and $\xi _{BSCO}\left(
0\right) =4$ nm. The results of the fits in these nonhysteretic Bi-2201
break-junctions are shown in Fig. 3 and the fitted values of the gap
depression are equal to $\left( x_0\right) _s=-4.0$ nm, $\left( x_0\right)
_d=0.5$ nm. The barrier thickness, as usual determined by the {\it s}-wave
depressed-gap fit, is $2w=9.8$ nm.

Fig. 4 reports data taken from a SIS\ YBCO-based bicrystal grain-boundary
junction fabricated on a SrTiO$_3$ (STO) substrate \cite{vale97} and fitted
by our model with a barrier thickness, determined as in the previous cases, $%
2w$ = $2.0$ nm . Here the amounts of gap depression are $\left( x_0\right)
_s $ = $-0.6$ nm and $\left( x_0\right) _d$ = $1.0$ nm, while $T_c=85$ K, $%
\Delta _{YBCO}\left( 0\right) =16$ meV and $\xi _{YBCO}\left( 0\right) =2.8$
nm.

Fig. 5 reports the case of a YBCO/STO/YBCO multilevel edge JJ \cite
{laibo94,gonn96a} with {\it ex situ} oxygen plasma treatment, which is well
fitted by our model (both {\it s}- and {\it d}-wave symmetry) with only one
adjustable parameter, being $2w=7.2$ nm already known independently from the
fabrication process. The fitted values of the parameter $x_0$ are $\left(
x_0\right) _s=-3.3$ nm and $\left( x_0\right) _d=-0.4$ nm. The physical
values characteristic of the superconducting state used in the fit are: $%
T_c=85$ K, $\Delta _{YBCO}\left( 0\right) =16$ meV and $\xi _{YBCO}\left(
0\right) =2.8$ nm. It is clear from Figs. 2 to 5 that both the {\it s}-wave
and the {\it d}-wave depressed-gap curves fit properly the magnitude and, in
most of the cases, also the shape of the various experimental $I_c(T)R_n(T) $
data, but with different amounts of gap depression:\ pure {\it d}-wave fit
requires a smaller amount of surface gap depression (i.e. a larger value of $%
x_0$) than pure {\it s}-wave one.

In the latter junction, taking advantage of the {\it a priori} knowledge of
the exact value of $w$, the behaviour of the amount of gap depression
(accounted for by $x_0$) vs. $\varepsilon $ (percentage of {\it d} wave) has
been obtained by using the general mixed $(s+id)$-wave model of Eq. (2) and
reported in Fig. 6. Since in that case $x_0^{dG}\simeq -1.9$ nm, from the
graph of Fig. 6 a pure {\it d}-wave would be excluded, and an upper limit of
the order of 40\% for the percentage of {\it d}-wave pair symmetry in YBCO
is draftily given. Anyway we must remark that the results are very sensitive
to the gap value and, therefore, different values for $\Delta _{YBCO}$ (such
as 20 meV rather than the 16 meV we used following the authors of Ref. \cite
{laibo94}) would yield for $x_0$ values compatible also with a pure {\it d}%
-wave pair symmetry. The $x_0^{dG}$ values, calculated by means of the above
mentioned approximate expression for the junctions of Fig. 2, 3 and 4, are
always comprised between $(x_0)_s$ and $(x_0)_d$. Even considering the
mentioned sensitivity of the results to the value of the gap and to the
value of $w$, this fact seems to indirectly question the presence of pure 
{\it d}-wave symmetry in the HTS materials of these junctions.

With this extension of our model, any comparison with the experimental
results of SIS' JJs \cite{sun94,france97} (where S' is a low-T$_c$
superconductor) is no longer meaningful due to the orthogonality between
wave functions with symmetries {\it s} and {\it d}. Actually, the mere
detection of a Josephson current in SIS' JJs rules out the existence of pure 
{\it d}-wave pair symmetry. Only depressed-gap pure {\it s}-wave symmetry 
\cite{gonn96a} or mixed ({\it s+id})-wave pair symmetry plus order-parameter
depression can be acceptable in SIS' JJs. In the latter case, if the ratio $%
\Delta _d/\Delta _s$ is not fixed (i.e., if it's a free parameter of the
model), its effect in reducing $I_c(T)R_n(T)$ values becomes not
distinguishable and not separable from the effect eventually operated by a
S-I gap depression. On the other hand, considering the ratio $\Delta
_d/\Delta _s$ as a free, adjustable parameter leads to an unphysical
sample-dependence of the pair symmetry which does not appear acceptable.

A first conclusion is therefore the full confirmation of the important role
played by a surface gap depression in reducing the ''quality factor'' $%
I_c(T)R_n(T)$ below its theoretical BCS value (regardless of the type of gap
symmetry considered) in HTS SIS Josephson junctions. As it is shown in Figs.
2 to 5 and in the framework of our model there is no possibility to fit the
shape and, especially, the magnitude of experimental data by using pure {\it %
s}-wave, pure {\it d}-wave or mixed ({\it s+id})-wave models without any gap
depression. Moreover, we remark that fitting of reduced $I_c\left( T\right)
/I_c\left( 0\right) $ values rather than $I_c\left( T\right) $ ones is not
meaningful - even if very often present in literature - since the main
discrepancy of experimental results from theoretical predictions is in the 
{\it magnitude} rather than in the {\it shape} of the curves.

As we have shown in the previous examples, the value of $x_0$ accounting for
the depression of the order parameter depends on the type of pair symmetry
adopted and, therefore, we can also conclude that - when applied to several
experimental results - this model might yield a contribution to the pair
symmetry debate only provided the amount of gap depression at S-I interface
in each case examined is given independently from other measurements or
physical considerations, and is not a free, adjustable parameter as we have
treated it in our present calculations.

FIG. 1. Surface-depressed order parameter $\Delta \left( x,T\right) $ as
function of $x$ and a ''slice'' $\delta \Delta _i$ into which we imagine to
section\ the gap, in order to compute $I_c(T)R_n(T)$ in SIS JJs.

FIG. 2. Comparison of the experimental $I_c(T)R_n(T)$ obtained from a Bi$_2$%
Sr$_2$CaCu$_2$O$_{8+x}$ single-crystal Josephson break junction with the
curves yielded by various theoretical models (see text): open symbols are
experimental data; dash-dots and dots represent predictions obtained
considering pure {\it s}-wave and pure {\it d}-wave pair symmetry without
gap depression, respectively, while the results of our model both in the
pure {\it s}-wave and in the pure {\it d}-wave cases are reported with a
dashed and continuous line, respectively.

FIG. 3. Comparison of the $I_c(T)R_n(T)$ behaviour obtained from a Bi$_2$Sr$%
_2$CuO$_{6+x}$ single-crystal break-junction with the results of various
models as shown in Fig. 2.

FIG. 4. The same as in Figs. 2 and 3 but with the $I_c(T)R_n(T)$ data taken
from a\ YBCO-based bicrystal grain-boundary junction fabricated on SrTiO$_3$
substrates (from Ref. \cite{vale97}).

FIG. 5. The same as in Figs. 2 - 4 but with the experimental $I_c(T)R_n(T)$
data taken from a SIS YBCO/STO/YBCO multilevel edge JJ (from Ref. \cite
{laibo94}) and  fitted by our model with only one adjustable parameter.

FIG. 6. Behaviour of the amount of gap depression (accounted for by $x_0$)
vs. percentage of {\it d} wave (in the general mixed $(s+id)$-wave model),
estimated taking advantage of  the {\it a priori} knowledge of the barrier
thickness $2w$ in the case of Fig. 5.

\end{document}